# PHOTONIC DRAG EFFECT FOR ONE-DIMENSIONAL ELECTRONS IN A LONGITUDINAL MAGNETIC FIELD WITH $D^{(-)}$-CENTERS PARTICIPATION

**V.D. Krevchik[1], A.B. Grunin[1], A.K. Aringazin[2,3], and M.B. Semenov[1,3]**
[1] Department of Physics, Penza State University, Penza, 440017, Russia
physics@diamond.stup.ac.ru
[2] Institute for Basic Research,
Eurasian National University, Astana, 473021, Kazakstan
aringazin@mail.kz
[3] Institute for Basic Research, P.O. Box 1577, Palm Harbor, Fl 34682, USA
ibr@gte.net

**Abstract**

The impurity photonic drag effect (PDE), theory for semiconductive quantum wire (QW) in a longitudinal (along the quantum wire axis) magnetic field $\vec{B}$, has been developed. The PDE is due to the photon longitudinal momentum transmission to localized electrons, under optical transitions from $D^{(-)}$-states to QW hybrid-quantum states, if the QW is described by the parabolic confinement potential. The analytical expression for the drag current (DC) density has been obtained within the framework of zero-range potential model and in the effective mass approximation. The drag current spectral dependence has been investigated for various values of $\vec{B}$ and QW parameters, under electron scattering on the dotty-impurities system. The drag current spectral dependence is characterized by Zeeman doublet with a pronounced "beak"-type peak. This peak is related to electron optical transitions from $D^{(-)}$-states to the states with the magnetic quantum number *m=1*. With an increase of the magnetic field $\vec{B}$ the beak-type peak is shifted to short-wave spectrum region, and the peak height considerably increases. We discuss the possibility of using of the one-dimensional drag current effect, in a longitudinal magnetic field, to develop a new type of laser radiation detectors.

## 1. Introduction

The photonic drag effect (PDE) is due to the photon momentum, which is transmitted to the electron (hole) subsystem during the absorption process. The account for the photon momentum leads to charge carriers distribution asymmetry in the quasi-momentum space, i.e. it leads to the appearance of the drag current (DC). The photonic drag effect for two-dimensional electrons at optical transitions between dimensionally quantized states for hetero-structures was theoretically investigated in [1]. As has been shown there, at certain conditions this effect can be sufficiently strong. The contribution of the intersubband and interband optical transitions to PDE for holes in the infinitely deep semi-conductive quantum well, have been considered in [2]. The $2D \rightarrow 1D$ dimensional reduction should lead to essential changes in physical properties of the quantum structures. In particular, a significant modification of local electron states and appearance of peculiarities in the light impurity absorption spectrum related to one-dimensional electron states specific character are expected. The governed modulation problem for binding energies of the impurity states [3], and, correspondingly, optical transitions energies control problem [4], stimulates studies on magneto-optical properties of quantum wires (QW). It has been shown in [5, 6] that magnetic field $\vec{B}$ applied along the QW axis can essentially change the lateral geometric confinement. Therefore, variations of $\vec{B}$ can affect the effective geometric size of the system, and hence can provide an ability to control optical properties of the system.

The aim of this work is to study the electron PDE, under QW $D^{(-)}$-centers photo-ionization in a longitudinal magnetic field. The PDE for one-dimensional electrons in this case is considered by the light with transversal polarization $\vec{e}_{\lambda\perp}$ (with respect to the QW axis) absorption, i.e. by the photon with the $\hbar\,\vec{q}_{||}$ momentum (along the QW axis) absorption.

To describe the QW one-electron states we can use the confinement parabolic potential $V\,(x,y) = m^{*}\,\omega_0^2\,(x^2 + y^2)/2$, where $m^{*}$ is the effective electron mass and $\omega_0$ is characteristic frequency for confinement potential. The vector-potential $\vec{A}(\vec{r})$ for constant uniform magnetic field can be chosen in the symmetric gauge, $\vec{A} = (-B\,y/2, B\,x/2, 0)$. Then, the effective hamiltonian for interaction with the light wave field, in cylindrical system of reference, can be written as

$$\hat{H}_{\text{int}} = -\lambda_0 \sqrt{\frac{2\pi\hbar^2\,\alpha^{*}\,I_0}{m^{*2}\,\omega}}\,\exp(i\,q_z\,z)\left[ i\,\hbar\left(\cos(\Theta-\varphi)\frac{\partial}{\partial\rho} + \frac{1}{\rho}\sin(\Theta-\varphi)\frac{\partial}{\partial\varphi}\right) + \right.$$

$$\left. + \frac{|e|B}{2}\,\rho\,\sin\,(\varphi-\Theta)\right], \qquad (1)$$

where $\rho, \varphi, z$ are cylindrical coordinates; $q_z$ is the photon wave vector projection, $\vec{q}_{||} = (0,0,q_z)$ to the QW axis; $\Theta$ is the light polarization vector $\vec{e}_{\lambda\perp}$ polar angle; $\lambda_0$ is the local field coefficient; $\alpha^{*}$ is the fine structure constant with account for the dielectric permeability $\varepsilon$; $I_0$ is the light intensity; $\omega$ is light frequency; $|e|$ is the electron charge; and $B$ is the magnetic induction.

The zero-range potential model can be used [7] for the impurity center (IC) potential $V_\delta(\vec{r}, \vec{R}_a)$:

$$V_\delta(\vec{\rho}, z, \vec{\rho}_a, z_a) = \gamma\,\delta(\vec{\rho} - \vec{\rho}_a)\,\delta(z - z_a)\left[1 + (\vec{\rho} - \vec{\rho}_a)\frac{\partial}{\partial\vec{\rho}} + (z - z_a)\frac{\partial}{\partial z}\right], \quad (2)$$

where $\gamma = 2\pi\hbar^2 / (\alpha\, m^*)$, and $\alpha$ can be determined by the binding energy $E_i$ for the electron localized state at the same impurity center in the massive semiconductor; IC is localized at the point $\vec{R}_a = (\vec{\rho}_a, z_a)$. As it is known [8], such a model can be used for $D^{(-)}$-states, corresponding to a description in terms of an additional electron joining to a small donor. As it will be shown below, the zero-range potential method allows one to obtain analytical solution for the localized carrier wave function, in an external longitudinal magnetic field. It is important for the positional disorder effect analysis (in QW with parabolic potential profile), and for the obtaining of the overt formula in the case of the one-dimensional electrons drag current. The energy spectrum in the chosen model has the following form [9]:

$$E_{n,m,k_z} = \frac{\hbar\,\omega_B\, m}{2} + \hbar\,\omega_0\sqrt{1 + \frac{\omega_B^2}{4\,\omega_0^2}}\,(2n + |m| + 1) + \frac{\hbar^2 k_z^2}{2m^*}, \quad (3)$$

where $n$ = 0, 1, 2, … is quantum number corresponding to Landau levels, $m$ = 0, ±1, ±2, … is magnetic quantum number, $\omega_B = |e|\,B / m^*$ is cyclotron frequency, and $\hbar k_z$ is the electron quasi-momentum projection to the z-axis.

The undisturbed (by impurities) one-electron states, $\Psi_{n,\,m,\,k_z}(\rho, \varphi, z)$, in the longitudinal magnetic field can be represented in the following form [9]:[1]

---

[1] Subsequently, we will consider the impurity electron strong localization case, i.e. $\lambda_B\, a_1$>>1, where $\lambda_B^{-1}$ is the localized state (in magnetic field) radius. This gives us grounds to assume that the one-electron states in longitudinal magnetic field have not been disturbed by the impurity potential.

$$\Psi_{n,\,m,\,k_z}(\rho,\varphi,z)=\frac{1}{2^{\frac{|m|+1}{2}}\sqrt{\pi L_{QW}}\,a_1^{|m|+1}}\left[\frac{n!}{(n+|m|)!}\right]^{\frac{1}{2}}\rho^{|m|}\exp\left(-\frac{\rho^2}{4a_1^2}\right)\times$$

$$\times L_n^{|m|}\left(\frac{\rho^2}{2a_1^2}\right)\exp(i\,m\,\varphi)\exp(i\,k_z\,z).$$

(4)

Here, $a_1^2=a^2/\left(2\sqrt{1+a^4/\left(4a_B^4\right)}\right)$, $a^2=\hbar/\left(m^*\omega_0\right)$; $a_B^2=\hbar/\left(m^*\omega_B\right)$, $L_n^{\alpha}(x)$ is Laguerre polynomial [10], and $L_{QW}$ is the QW length.

In this paper, one-electrons drag current (DC) under the IC photoionization for a strong magnetic quantization case, when the oscillator characteristic length considerably exceeds the magnetic length, has been calculated. The elastic scattering of electron off the short-range impurities system, which has been modeled by the zero-range potential sum, has been accounted [11-13].

## 2. The $D^{(-)}$-center binding energy in a longitudinal magnetic field

Let us consider the positional disorder effect for semiconductive QW with the confining parabolic potential in a longitudinal magnetic field. It is supposed that the impurity center is situated at point $\vec{R}_a = (\rho_a, \varphi_a, z_a)$. The wave function for electron, which is localized at the $D^{(-)}$-center, satisfies Lippman-Schwinger equation for a bound state. In cylindrical system of reference, this equation is written as

$$\Psi_{\lambda_B}(\rho,\varphi,z;\rho_a,\varphi_a,z_a)=\int_0^{\infty}\int_0^{2\pi}\int_{-\infty}^{+\infty}\rho_1\,d\rho_1\,d\varphi_1\,dz_1\,G\left(\rho,\varphi,z,\rho_1,\varphi_1,z_1;E_{\lambda_B}^{(0)}\right)\times$$
$$\times V_\delta(\rho_1,\varphi_1,z_1;\rho_a,\varphi_a,z_a)\Psi_{\lambda_B}(\rho_1,\varphi_1,z_1;\rho_a,\varphi_a,z_a),$$

(5)

where $G\left(\rho,\varphi,z,\rho_1,\varphi_1,z_1;E_{\lambda_B}^{(0)}\right)$ is the one-electron Green function, which corresponds to a source at the point $(\rho_1,\varphi_1,z_1)$ and to the energy $E_{\lambda_B}^{(0)}=-\hbar^2\lambda_B^2/(2m^*)$ ($E_{\lambda_B}^{(0)}$ are eigenvalues of the hamiltonian $\hat{H}_B=\left(\hat{\vec{p}}+|e|\vec{A}\right)^2/(2m^*)+V(x,y)+V_\delta(\vec{\rho},z,\vec{\rho}_a,z_a)$, where $\hat{\vec{p}}$ is the electron momentum operator):

$$G\left(\rho,\varphi,z,\rho_1,\varphi_1,z_1;E_{\lambda_B}^{(0)}\right)=\int_{-\infty}^{+\infty}d\left(\frac{k_z L_{QW}}{2\pi}\right)\sum_{n,m}\frac{\Psi_{n,m,k_z}^{*}(\rho_1,\varphi_1,z_1)\Psi_{n,m,k_z}(\rho,\varphi,z)}{E_{\lambda_B}^{(0)}-E_{n,m,k_z}}$$

(6)

Substituting Eq. (2) into Eq. (5) gives

$$\Psi_{\lambda_B}(\rho,\varphi,z;\rho_a,\varphi_a,z_a)=\gamma G\left(\rho,\varphi,z,\rho_a,\varphi_a,z_a;E^{(0)}_{\lambda_B}\right)\times$$
$$\times\left(\hat{T}\,\Psi_{\lambda_B}\right)(\rho_a,\varphi_a,z_a;\rho_a,\varphi_a,z_a), \qquad (7)$$

where

$$\left(\hat{T}\,\Psi_{\lambda_B}\right)(\rho_a,\varphi_a,z_a;\rho_a,\varphi_a,z_a)\equiv$$
$$\equiv \lim_{\substack{\rho\to\rho_a\\ \varphi\to\varphi_a\\ z\to z_a}}\left[1+(\vec{\rho}-\vec{\rho}_a)\frac{\partial}{\partial\vec{\rho}}+(z-z_a)\frac{\partial}{\partial z}\right]\Psi_{\lambda_B}(\rho,\varphi,z;\rho_a,\varphi_a,z_a) \qquad (8)$$

Application of Eq. (8) to both sides of Eq. (7) gives us the equation, which determines the IC binding energy $E^{(0)}_{\lambda_B}$ dependence on QW parameters, IC coordinates and magnetic induction $B$,

$$1=\gamma\left(\hat{T}\,G\right)\left(\rho_a,\varphi_a,z_a,\rho_a,\varphi_a,z_a;E^{(0)}_{\lambda_B}\right) \qquad (9)$$

Due to Eq. (9), the electron bound state energy in the total field $E^{(0)}_{\lambda_B}$ is the pole of the Green function. According to Eqs. (3), (4), and (6), this Green function can be written as

$$G\left(\rho,\varphi,z,\rho_a,\varphi_a,z_a;E^{(0)}_{\lambda_B}\right)=-\frac{\hbar^2}{4\pi^2\,m^*\,E_d\,a_1^2}\int_{-\infty}^{+\infty}dk_z\,\exp\left[i\,k_z\,(z-z_a)\right]\times$$
$$\times\sum_{n,\,m}C^2_{n,m}\,(\rho_a\,\rho)^{|m|}\exp\left(-\frac{\rho_a^2+\rho^2}{4a_1^2}\right)L_n^{|m|}\left(\frac{\rho_a^2}{2a_1^2}\right)L_n^{|m|}\left(\frac{\rho^2}{2a_1^2}\right)\times$$
$$\times\exp\left[i\,m(\varphi-\varphi_a)\right]\left(\eta_B^2+m\,a^{*\,-2}+\beta^{-1}\sqrt{1+\beta^2\,a^{*\,-4}}\,(2n+|m|+1)+k_z^2a_d^2\right)^{-1}, \qquad (10)$$

where $C_{n,m} = 2^{-\frac{|m|}{2}} a_1^{-|m|} \sqrt{n!/(n+|m|)!}$; $a^* = a_B / a_d$; $a_d$ is the effective Bohr radius; $\eta_B^2 = \left|E_{\lambda_B}^{(0)}\right| / E_d$; $E_d$ is the effective Bohr energy; $\beta = L^* / \left(4\sqrt{U_0^*}\right)$; $L^* = 2L/a_d$; $2L$ is QW diameter; $U_0^* = U_0 / E_d$; $U_0$ is QW potential amplitude.

The summation in Eq. (10) over $n$ can be fulfilled with the help of the relation,

$$\left(\eta_B^2 + m a^{*^{-2}} + \beta^{-1}\sqrt{1+\beta^2 a^{*^{-4}}}\,(2n+|m|+1) + k_z^2 a_d^2\right)^{-1} =$$

$$= \int_0^{+\infty} \exp\left[-\left(\eta_B^2 + m a^{*^{-2}} + \beta^{-1}\sqrt{1+\beta^2 a^{*^{-4}}}\,(2n+|m|+1) + k_z^2 a_d^2\right)t\right] dt,$$

(11)

and Hille-Hardi formula for bilinear generating function [10],

$$\sum_{n=0}^{\infty} \frac{n!}{\Gamma(n+\alpha+1)} L_n^{\alpha}(x) L_n^{\alpha}(y) z^n = (1-z)^{-1} \exp\left(-z\frac{x+y}{1-z}\right)(xyz)^{-\frac{\alpha}{2}} I_\alpha\left(2\frac{\sqrt{xyz}}{1-z}\right)$$

(12)

Here $|z|<1$, $I_\alpha(u)$ is the modified Bessel function of first kind [10]. Then, the series sum in Eq. (10) can be performed over *m*, with the use of the generating function of Bessel functions $J_k(z)$ of first kind [10],

$$\sum_{k=-\infty}^{+\infty} J_k(z) t^k = \exp\left[\frac{1}{2}\left(t - \frac{1}{t}\right)z\right].$$

(13)

Finally, taking into account that performing the integral over $k_z$ in Eq. (10) gives

$$\int_{-\infty}^{+\infty} \exp\left[i k_z \left(z - z_a\right) - k_z^2 a_d^2 t\right] dk_z = \frac{1}{a_d}\sqrt{\frac{\pi}{t}} \exp\left[-\frac{\left(z - z_a\right)^2}{4 a_d^2 t}\right], \tag{14}$$

with the help of Weber integral [10] (in the notation used in this paper),

$$\int_0^{+\infty} \frac{1}{t\sqrt{t}} \exp\left[-\frac{\left(\rho - \rho_a\right)^2}{4 a_1^2 t} - \frac{\left(z - z_a\right)^2}{2 a^2 t}\right] \exp\left[-\left(\beta\eta_B^2 + \sqrt{1 + \beta^2 a^{*-4}}\right) t\right] dt =$$

$$= \sqrt{2\pi}\, a \frac{\exp\left[-\sqrt{2\left(\beta\eta_B^2 + \sqrt{1+\beta^2 a^{*-4}}\right)} \sqrt{\frac{\left(\rho - \rho_a\right)^2 \sqrt{1 + \beta^2 a^{*-4}} + \left(z - z_a\right)^2}{a^2}}\right]}{\sqrt{\left(\rho - \rho_a\right)^2 \sqrt{1 + \beta^2 a^{*-4}} + \left(z - z_a\right)^2}} \tag{15}$$

the one-electron Green function in Eq. (10) can be represented as

$$G\left(\rho, \varphi, z, \rho_a, \varphi_a, z_a; E_{\lambda_B}^{(0)}\right) = -\frac{\hbar^2}{2^3 \pi^{\frac{3}{2}} E_d\, a_d^3\, m^* \sqrt{\beta}} \left\{ \int_0^{+\infty} \frac{1}{\sqrt{t}} \times \right.$$

$$\times \exp\left[-\left(\left(\beta\eta_B^2 + w\right)t + \frac{\left(z - z_a\right)^2}{4\beta a_d^2 t}\right)\right] \times \left[2w\left(1 - \exp(-2wt)\right)^{-1} \exp\left[-\frac{\left(\rho_a^2 + \rho^2\right) w}{4\beta a_d^2} \frac{\left(1 + \exp(-2wt)\right)}{\left(1 - \exp(-2wt)\right)}\right] \times\right.$$

$$\times \exp\left[\frac{1}{2}\left(\exp\left(i\left(\varphi - \varphi_a\right) - \frac{\beta}{a^{*2}} t\right) + \exp\left(-i\left(\varphi - \varphi_a\right) + \frac{\beta}{a^{*2}} t\right)\right) \frac{\rho_a \rho\, w \exp(-wt)}{\beta a_d^2 \left(1 - \exp(-2wt)\right)}\right] -$$

$$-\frac{1}{t}\exp\left[-\frac{(\rho-\rho_a)^2 w}{4\beta a_d^2 t}\right]\Bigg]dt+2\sqrt{\pi\beta}a_d\frac{\exp\left[-\frac{1}{a_d}\sqrt{\frac{\beta\eta_B^2+w}{\beta}}\sqrt{(\rho-\rho_a)^2 w+(z-z_a)^2}\right]}{\sqrt{(\rho-\rho_a)^2 w+(z-z_a)^2}}\Bigg\},$$

(16)

where $w=\sqrt{1+\beta^2 a^{*-4}}$.

Substituting Eq. (16) into Eq. (9) and taking the limits gives us the equation for a determination of the $D^{(-)}$ -center binding energy, in longitudinal magnetic field (in Bohr units),

$$\sqrt{\eta_B^2+\beta^{-1}w}=\eta_i-\frac{1}{\sqrt{\pi\beta}}\int_0^{+\infty}\frac{1}{\sqrt{t}}\exp\left[-\left(\beta\eta_B^2+w\right)t\right]\left(\frac{1}{2t}-w\left(1-\exp(-2wt)\right)^{-1}\times\right.$$
$$\left.\times\exp\left[-\frac{\rho_a^{*2}w}{2\beta(1-\exp(-2wt))}\left(1+\exp(-2wt)-\left(\exp\left(-\beta a^{*-2}t\right)+\exp\left(\beta a^{*-2}t\right)\right)\exp(-wt)\right)\right]\right)dt,$$

(17)

where $\rho_a^*=\rho_a/a_d$.

Equation (17) can be analyzed numerically by using computer. However, it is necessary to account for following two circumstances. First, the localized states can be situated between the QW bottom and the first dimensionally quantized level $\varepsilon_{0,0}$ [8]. In this case, for impurity levels, which are situated at higher positions than the QW bottom; $E_{\lambda_B}^{(0)}>0$ and the $\lambda_B$ parameter becomes imaginary. Second, because of the quantum dimensional effect, the binding energy of $D^{(-)}$ -center $E_{\lambda_B}$ of QW in longitudinal magnetic field should be determined as [14]

$$E_{\lambda_B} = \begin{cases} \varepsilon_{0,0} + \left| E_{\lambda_B}^{(0)} \right|, \ E_{\lambda_B}^{(0)} < 0, \\ \varepsilon_{0,0} - E_{\lambda_B}^{(0)}, \ E_{\lambda_B}^{(0)} > 0 \end{cases}$$

or, in Bohr units,

$$E_{\lambda_B} / E_d = \begin{cases} \beta^{-1} w + \eta_B^2, \ E_{\lambda_B}^{(0)} < 0, \\ \beta^{-1} w - \eta_B'^{\,2}, \ E_{\lambda_B}^{(0)} > 0, \end{cases} \qquad (18)$$

where $\varepsilon_{0,0} = \hbar\omega_0 \sqrt{1 + \omega_B^2 / \left(4\omega_0^2\right)}$; $\eta_B'^{\,2} = -\eta_B^2$.

In Fig. 1, we present the result of the numerical analysis of semiconductive QW $D^{(-)}$-states (based on InSb) of Eq. (17) with the account for Eq. (18); the effective mass of electron in InSb and dielectric permeability are: $m^* = 0.0133 m_0$ (where $m_0$ is the electron mass at rest) and $\varepsilon \approx 18$, correspondingly, and the effective Bohr energy is $E_d \approx 5.5 \times 10^{-4}$ eV. As one can see from Fig. 1, in both the cases, $E_{\lambda_B}^{(0)}$>0 and $E_{\lambda_B}^{(0)}$<0 (curves 1 and 2, respectively), the binding energy of $D^{(-)}$-center $E_{\lambda_B}$ is a decreasing function of its ($D^{(-)}$-center) radial coordinate $\rho_a$, that is related to an essential modification of the local electron states near the QW boundaries. The $D^{(-)}$-center binding energy considerably increases in the presence of magnetic field (see curves 3 and 4 of Fig. 1). In the case $E_{\lambda_B}^{(0)}$<0, the binding energy increases, as one can see from Fig. 2 (see, for example, curve 2), by more than 0.02 eV for the $D^{(-)}$-center, which is situated at the origin of coordinates. Then, the bound state existence condition, in longitudinal magnetic field, becomes less restrictive, as one can see by comparing curves 1 and 3, 2 and 4 of Fig. 1. Hence, magnetic field stabilizes the QW $D^{(-)}$ -state. It should be noted that the enhancement of $D^{(-)}$ -centers binding energy with an increase of the magnetic field $\vec{B}$ is experimentally observed in the quantum multi-well systems of GaAs-$Ga_{0.75}$ Al $_{0.25}$As [15]. The arising possibility of effective control of optical transitions energies in magnetic field is of interest. This could allow to construct photoreceivers with variable working frequency and with sensitivity, in the light impurity absorption region.

## 3. The drag current calculation for one-dimensional electrons in longitudinal magnetic field

As it follows from Eq. (7), the wave function for electron, which is localized on the short-range potential, $\Psi_{\lambda_B}(\rho, \varphi, z; \rho_a, \varphi_a, z_a)$, differs from the one-electron Green function $G\left(\rho, \varphi, z, \rho_a, \varphi_a, z_a; E_{\lambda_B}^{(0)}\right)$ only by the factor,

$$\Psi_{\lambda_B}(\rho, \varphi, z; \rho_a, \varphi_a, z_a) = -C\,\tilde{G}\left(\rho, \varphi, z, \rho_a, \varphi_a, z_a; E_{\lambda_B}^{(0)}\right), \qquad (19)$$

where

$$\tilde{G}\left(\rho, \varphi, z, \rho_a, \varphi_a, z_a; E_{\lambda_B}^{(0)}\right) = 2^{3/2}\beta^{1/2} E_d\, a_d^3\, G\left(\rho, \varphi, z, \rho_a, \varphi_a, z_a; E_{\lambda_B}^{(0)}\right);$$

$$C = \left[2^{3/2}\beta^{1/2} a_d^3 \frac{\partial \tilde{G}}{\partial\left(\eta_B^2\right)}\left(\rho_a, \varphi_a, z_a, \rho_a, \varphi_a, z_a; E_{\lambda_B}^{(0)}\right)\right]^{-1/2}$$

is the normalization factor. For the case when $D^{(-)}$ -center is localized in the point $\vec{R}_a = (0, 0, z_a)$, from Eq. (10) we obtain

$$\Psi_{\lambda_B}(\rho,\varphi,z;0,0,z_a)\equiv\Psi_{\lambda_B}(\rho,\varphi,z;z_a)=2^{1/4}\pi^{-1/4}\beta^{-3/4}a_d^{-3/2}w^{5/4}\times$$

$$\times\left[\zeta\left(\frac{3}{2},\frac{\beta\eta_B^2}{2w}+\frac{1}{2}\right)\right]^{-1/2}\times$$

$$\times\int_0^\infty\frac{1}{\sqrt{t}}\exp\left[-\left((\beta\eta_B^2+w)t+\frac{(z-z_a)^2}{4\beta a_d^2 t}\right)\right]\times[1-\exp(-2wt)]^{-1}\times$$

$$\times\exp\left[-\frac{\rho^2 w(1+\exp(-2wt))}{4\beta a_d^2(1-\exp(-2wt))}\right]dt, \quad (20)$$

where $\zeta(s,u)$ is generalized Riemann zeta-function [10].

The impurity photon drag effect (PDE) in the solution of the quantum wire (QW) problem is based on the Boltzmann kinetic equation, which is written in the relaxation time approximation. The generative term of this equation is determined by quantum photo-transitions of carriers from the $D^{(-)}$ -center to the hybrid-quantizing band. These terms can be calculated in the linear (with respect to photon momentum) approximation. In the short-circuit regime conditions the electron drag current density $j(\omega)$ for QW in a longitudinal magnetic field has the following form:

$$j(\omega)=-\frac{|e|N_0}{2\pi^2\hbar^2}\int_0^{L_{QW}}dz_a\cdot n_\lambda\cdot\sum_{n,m}\theta\left[\hbar\omega-\left|E_{\lambda_B}^{(0)}\right|-\frac{\hbar\omega_B m}{2}-\hbar\omega_0\sqrt{1+\frac{\omega_B^2}{4\omega_0^2}}(2n+|m|+1)\right]\times$$

$$\times\int_{-\infty}^{+\infty}\frac{\partial E_{n,m,k_z}}{\partial k_z}\cdot\tau\left(E_{n,m,k_z}\right)\cdot\left|M_{f,\lambda}\right|^2\cdot\left[f_0\left(E_{\lambda_B}^{(0)}\right)-f_0\left(E_{n,m,k_z}\right)\right]\times$$

$$\times\delta\left[\hbar\omega-\left|E_{\lambda_B}^{(0)}\right|-\frac{\hbar\omega_B m}{2}-\hbar\omega_0\sqrt{1+\frac{\omega_B^2}{4\omega_0^2}}(2n+|m|+1)-\frac{\hbar^2k_z^2}{2m^*}\right]dk_z, \quad (21)$$

where $N_0$ is the QW $D^{(-)}$ -centers concentration; $n_\lambda$ is the $D^{(-)}$ -centers linear concentration, (for $D^{(-)}$ -centers, which are localized in the points $\vec{R}_a = (0, 0, z_a)$ of the QW axis); $\hbar\omega$ is photon energy; $\tau\left(E_{n,m,k_z}\right)$ is relaxation time for the QW electrons; $f_0(E)$ is the quasi-equilibrium distribution function for the QW electrons; $\delta(x)$ is Dirac delta-function; $M_{f,\lambda}$ are matrix elements, which determine the electron optical transitions from $D^{(-)}$ -center ground state to hybrid-quantizing QW states; $\theta(s) = \begin{cases} 1, \text{ïðè} \;\; s \ge 0, \\ 0, \text{ïðè} \;\; s < 0 \end{cases}$ is Heaviside unit-step function [16].

With the account for Eq. (1) matrix elements $M_{f,\lambda}$ can be written as the sum of two components, $M_{f,\lambda} = I_1 + I_2$, where

$$I_1 = -\lambda_0 \sqrt{\frac{2\pi\hbar^2\alpha^*}{m^{*2}\omega}} I_0 \times$$

$$\times \left\langle \Psi^*_{n,m,k_z}(\rho,\varphi,z) \right| e^{i q_z z} \cdot i\hbar \cdot \left( \cos(\Theta-\varphi)\frac{\partial}{\partial\rho} + \frac{1}{\rho}\sin(\Theta-\varphi)\frac{\partial}{\partial\varphi} \right) \left| \Psi_{\lambda_B}(\rho,\varphi,z;z_a) \right\rangle, \tag{22}$$

$$I_2 = -\lambda_0 \sqrt{\frac{2\pi\hbar^2\alpha^*}{m^{*2}\omega}} I_0 \left\langle \Psi^*_{n,m,k_z}(\rho,\varphi,z) \right| e^{i q_z z} \cdot \frac{|e|B}{2} \cdot \rho \cdot \sin(\varphi-\Theta) \left| \Psi_{\lambda_B}(\rho,\varphi,z;z_a) \right\rangle \tag{23}$$

Calculation of the above $I_1$ requires the use of the following integrals:

$$\int_0^{2\pi} \exp(-i\,m\varphi)\cos(\varphi-\Theta)\,d\varphi = \begin{cases} \pi \cdot \exp(\mp i\Theta), & if \;\; m = \pm 1, \\ 0, & if \;\; m \ne \pm 1, \end{cases} \tag{24}$$

$$\int_{-\infty}^{+\infty} \exp\left[i(q_z-k_z)z-\frac{(z-z_a)^2}{4\beta a_d^2 t}\right]dz=2a_d\sqrt{\pi\beta}\sqrt{t}\exp\left[-\beta a_d^2 (q_z-k_z)^2 t+i(q_z-k_z)z_a\right]. \tag{25}$$

Performing the integral in Eq. (22) over $\rho$, with the account for known relation [10],

$$\int_0^{+\infty} \rho^3 \exp\left[-\frac{\rho^2 w}{2\beta a_d^2 (1-\exp(-2wt))}\right] L_n^1\left(\frac{\rho^2 w}{2\beta a_d^2}\right) d\rho =$$

$$=\frac{2\beta^2 a_d^4}{w^2}(n+1)(1-\exp(-2wt))^2 \exp(-2nwt), \tag{26}$$

yields the following result:

$$I_1 = 2^{11/4} i \pi^{1/2} L_{QW}^{-1/2} \exp(\mp i\Theta)\lambda_0 \sqrt{\frac{\alpha^* I_0}{\omega}}\, \beta^{-1/4} E_d a_d^{3/2} w^{5/4} (n+1)^{1/2} \exp[i(q_z-k_z)z_a]\times$$

$$\times\left[\zeta\left(\frac{3}{2},\frac{\beta\eta_B^2}{2w}+\frac{1}{2}\right)\right]^{-1/2} \left[\beta\eta_B^2+(2n+2)w+\beta a_d^2 (q_z-k_z)^2\right]\times$$

$$\times\left[\beta\eta_B^2+(2n+1)w+\beta a_d^2 (q_z-k_z)^2\right]^{-1}\cdot\left[\beta\eta_B^2+(2n+3)w+\beta a_d^2 (q_z-k_z)^2\right]^{-1}. \tag{27}$$

In the case of $I_2$ the integrals over $z$ and $\rho$ coordinates coincide with (25) and (26), correspondingly while in integrating over $\varphi$ we use

$$\int_0^{2\pi} \exp(-im\varphi)\sin(\varphi-\Theta)d\varphi=\begin{cases}\mp\pi i \exp(\mp i\Theta), & if\ m=\pm 1,\\ 0, & if\ m\neq\pm 1\end{cases}. \tag{28}$$

As it follows from Eqs. (24) and (28), optical transitions from impurity level are possible only to the states with quantum number values $m=\pm 1$. Accounting for Eqs. (25), (26), and (28) for $I_2$ we obtain

$$I_2 = \pm 2^{11/4} i \pi^{1/2} L_{QW}^{-1/2} \exp(\mp i\Theta)\lambda_0 \sqrt{\frac{\alpha^* I_0}{\omega}} \beta^{3/4} a^{*-2} E_d a_d^{3/2} w^{5/4} (n+1)^{1/2} \times$$

$$\times \exp[i(q_z - k_z) z_a] \times \left[\zeta\left(\frac{3}{2}, \frac{\beta\eta_B^2}{2w} + \frac{1}{2}\right)\right]^{-1/2} \times$$

$$\times \left[\beta\eta_B^2 + (2n+1)w + \beta a_d^2 (q_z - k_z)^2\right]^{-1} \cdot \left[\beta\eta_B^2 + (2n+3)w + \beta a_d^2 (q_z - k_z)^2\right]^{-1} \tag{29}$$

In the linear approximation (in $q_z$ ) for $\left|M_{f,\lambda}\right|^2$, entering Eq. (21), we obtain

$$\left|M_{f,\lambda}\right|^2 = \left|I_1 + I_2\right|^2 = \frac{2^{15/2}\pi}{L_{QW}} \lambda_0^2 \frac{\alpha^* I_0}{\omega} a_d^5 E_d^2 \beta^{1/2} w^{5/2} q_z \left[\zeta\left(\frac{3}{2}, \frac{\beta\eta_B^2}{2w} + \frac{1}{2}\right)\right]^{-1} (n+1) k_z \times$$

$$\times \left[\frac{2\left(\beta\eta_B^2 + (2n+2)w + \beta a_d^2 k_z^2\right)}{\left(\beta\eta_B^2 + (2n+1)w + \beta a_d^2 k_z^2\right)\left(\beta\eta_B^2 + (2n+3)w + \beta a_d^2 k_z^2\right)} - \right.$$

$$\left. - \left(\beta\eta_B^2 + (2n+2)w \pm \beta a^{*-2} + \beta a_d^2 k_z^2\right)^{-1}\right] \times$$

$$\times \frac{\left(\beta\eta_B^2 + (2n+2)w \pm \beta a^{*-2} + \beta a_d^2 k_z^2\right)^2}{\left(\beta\eta_B^2 + (2n+1)w + \beta a_d^2 k_z^2\right)^2 \left(\beta\eta_B^2 + (2n+3)w + \beta a_d^2 k_z^2\right)^2} . \tag{30}$$

Substituting Eq. (30) into Eq. (21), for the drag current density we have

$$j(\omega) = -\frac{|e| q_z N_0}{\pi \hbar L_{QW}} \cdot 2^{\frac{15}{2}} \lambda_0^2 \alpha^* I_0 n_\lambda a_d^4 E_d \beta^{\frac{5}{2}} w^{\frac{5}{2}} X \times$$

$$\times \int_0^{L_{QW}} dz_a \left[\zeta\left(\frac{3}{2}, \frac{\beta\eta_B^2}{2w} + \frac{1}{2}\right)\right]^{-1} \tau\left(E_d \left(X - \eta_B^2\right)\right)\left[f_0\left(-E_d \eta_B^2\right) - f_0\left(E_d \left(X - \eta_B^2\right)\right)\right] \times$$

$$\times \sum_{n=0}^{N} (n+1) \sum_{m=-1}^{1} \delta_{|m|,1}\, \theta\left( X - \eta_B^2 - \frac{m}{a^{*2}} - \frac{w}{\beta}\left(2n+|m|+1\right) \right)\times$$

$$\times \frac{\sqrt{X - \eta_B^2 - \frac{m}{a^{*2}} - \frac{w}{\beta}\left(2n+|m|+1\right)}}{\left(\beta\left(X - m a^{*-2}\right) - |m| w\right)^2 \left(\beta\left(X - m a^{*-2}\right) + \left(2-|m|\right)w\right)^2}\times$$

$$\times\left[\frac{2\beta\left(X - m a^{*-2}\right)}{\left(\beta\left(X-m a^{*-2}\right)-|m| w\right)\left(\beta\left(X-m a^{*-2}\right)+\left(2-|m|\right)w\right)} - \beta^{-1}X^{-1}\right],$$

(31)

where $X = \hbar\omega / E_d$ is photon energy in the effective Bohr energy $E_d$ units; $N = \lfloor A_0 \rfloor$ is an integer part of the number $A_0 = \beta\left(X - \eta_B^2 + a^{*-2}\right)/(2w) - 1$; $\delta_{|m|,1} = \begin{cases} 1, \text{åñëè } m = \pm 1, \\ 0, \text{åñëè } m \neq \pm 1 \end{cases}$ is Kronecker symbol, which accounts for the selection rule for the magnetic quantum number $m$.

In obtaining Eq. (31) for the drag current density we have assumed that integrating over $k_z$ in Eq. (21) requires to calculate roots of Dirac delta-function, $(k_z)_{1,2}$, which satisfy

$$X - \eta_B^2 - \frac{m}{a^{*2}} - \frac{w}{\beta}\left(2n+|m|+1\right) - k_z^2\, a_d^2 = 0. \qquad (32)$$

To study the drag current density spectral dependence, one has to investigate specific mechanism of scattering of the charge carriers (in QW). Accordingly, the relaxation time, in Eq. (31), has to be determined. We assume that the short-range impurities potentials [17] cause an elastic scattering for electrons in the QW hybrid-quantizing conductance band. Then, in the strong mag-netic quantization approximation, i.e. for $a_1 << a$, the relaxation time, $\tau\left(E_d\left(X-\eta_B^2\right)\right)$, can be written as [17]

$$\tau\left(E_d\left(X-\eta_B^2\right)\right)=2^{-\frac{9}{2}}\pi^{-1}\hbar E_d^{-1}L^{*2}\left(n_i a_d^3\right)^{-1}\left(\frac{\lambda_s}{a_d}\right)^{-2}\sqrt{\frac{w}{\beta}}\times$$

$$\times\left|1+\frac{1}{\sqrt{2}a^*}\left(\frac{\lambda_s}{a_d}\right)\zeta\left(\frac{1}{2},\frac{1}{2}-\frac{\beta\left(X-\eta_B^2\right)}{2w}\right)\right|^2\left[\sum_{n=0}^{N_1}\left(\frac{\beta\left(X-\eta_B^2\right)}{2w}-n-\frac{1}{2}\right)^{-1/2}\right]^{-1}$$

(33)

where $\lambda_s$ is the scattering length; $n_i$ is the impurity scattering centers concentration in QW; $N_1=\lfloor A_1\rfloor$ is an integer part of $A_1$; $A_1=\beta\left(X-\eta_B^2\right)/(2w)-1/2$, if $\lfloor A_1\rfloor\neq A_1$ and $N_1=\lfloor A_1\rfloor-1$, if $\lfloor A_1\rfloor=A_1$.

Due to the work [17], QW electrons distribution function, $f_0\left(E_{n,\,m,\,k_z}\right)$, in the considered case can be represented as

$$f_0\left(E_{n,\,m,\,k_z}\right)=8\sqrt{\pi}\left(n_e a_d^3\right)\sqrt{\delta_T}\,\beta w^{-1}sh\left(\delta_T w\beta^{-1}\right)\exp\left(-\delta_T\frac{E_{n,\,m,\,k_z}}{E_d}\right),$$

(34)

where $n_e$ is concentration of electrons, $\delta_T=E_d/(kT)$, $T$ is thermodynamical temperature, and $k$ is Boltzmann constant.

Let us consider the case $\delta_T=1$. For example, for QW, based on InSb, this corresponds to $T\approx 7K$, and impurity centers can be considered as fully occupied, i.e. in Eq. (31) we can use $f_0\left(-E_d\,\eta_B^2\right)=1$. Then, Eq. (31) for the drag current density, with account for Eqs. (33) and (34), becomes

$$j(\omega)= j_0\, L^{*2}\, n_\lambda\, a_d \left(n_i\, a_d^3\right)^{-1}\left(\frac{\lambda_s}{a_d}\right)^{-2}\beta^2\, w^3\, X\left[\zeta\left(\frac{3}{2},\frac{\beta\eta_B^2}{2w}+\frac{1}{2}\right)\right]^{-1}\times$$

$$\times\left|1+\frac{1}{\sqrt{2}\,a^*}\left(\frac{\lambda_s}{a_d}\right)\zeta\left(\frac{1}{2},\frac{1}{2}-\frac{\beta\left(X-\eta_B^2\right)}{2w}\right)\right|^2\left[\sum_{n=0}^{N_1}\left(\frac{\beta\left(X-\eta_B^2\right)}{2w}-n-\frac{1}{2}\right)^{-1/2}\right]^{-1}\times$$

$$\times\left[1-8\sqrt{\pi}\left(n_e\, a_d^3\right)\sqrt{\delta_T}\,\beta\, w^{-1}\, sh\left(\delta_T\, w\beta^{-1}\right)\exp\left[-\delta_T\left(X-\eta_B^2\right)\right]\right]\times$$

$$\times\sum_{n=0}^{N}(n+1)\sum_{m=-1}^{1}\delta_{|m|,1}\,\theta\left(X-\eta_B^2-\frac{m}{a^{*2}}-\frac{w}{\beta}\left(2n+|m|+1\right)\right)\times$$

$$\times\frac{\sqrt{X-\eta_B^2-\dfrac{m}{a^{*2}}-\dfrac{w}{\beta}\left(2n+|m|+1\right)}}{\left(\beta\left(X-m\,a^{*-2}\right)-|m|\,w\right)^2\left(\beta\left(X-m\,a^{*-2}\right)+\left(2-|m|\right)w\right)^2}\times$$

$$\times\left[\frac{2\beta\left(X-m\,a^{*-2}\right)}{\left(\beta\left(X-m\,a^{*-2}\right)-|m|\,w\right)\left(\beta\left(X-m\,a^{*-2}\right)+\left(2-|m|\right)w\right)}-\beta^{-1}X^{-1}\right],$$

(35)

where $j_0=-4\pi^{-3}\lambda_0^2\,\alpha^*\,N_0\,|e|\,a_d^4\,I_0\,q_z$ .

Using the numerical values $E_i=0.06\ eV$, $L\cong 43$ *nm*, $n_i=2.7\times10^{15}\ cm^{-3}$, $n_e=1.4\times10^{16}\ cm^{-3}$, $U_0=0.2\ eV$, $\lambda_s\cong 29$ *nm*, $\hbar v=0.21\ eV$, and $B=10\ T$ in Eq. (35), we obtain the estimation of the drag current density for QW based on InSb, $j(\omega)\cong(1.7\times10^{-18}\ N_0)\ A/m^2$ . For $N_0 = 10^{15}$ $cm^{-3}$ we have $j(\omega) \cong 1.7 \times 10^{-1}$ $A/cm^2$ , i.e. we predict an enhancement by more than one order of magnitude relative to the corresponding value in a single semiconductive quantum well in zero magnetic field [18]. Fig. 3 shows the drag current density spectral dependence for one-dimensional electrons (in relative units $j / j_0$ ) under $D^{(-)}$ -centers photo-ionization in longitudinal magnetic field. As one can see from Fig. 3, the drag current density spectral dependence is characterized by the Zeeman doublet with pronounced "beak" type peak. This peak is related to electrons optical transitions from $D^{(-)}$ -states to states with $m = 1$ (the magnetic quantum number value). With the increase of the field $\vec{B}$ the "beak" type peak is shifted to the short-wave spectrum region, and the peak height considerably increases (compare curves 1 and 2). Also, Fig. 3 shows that the variation of the magnetic field by 2 *T* (corresponding to the beak-type peak under transition from curve 1 to curve 2), leads to the wave length decrease approximately by 1 *mkm* . It allows us to conclude that it is possible to construct a photo-receiver (photo-sensor) based on the drag current effect in QW semiconductive structures, with variable sensitivity in magnetic field.

## 4. Conclusions

A theoretical study of the impurity drag current effect in longitudinal magnetic field developed in this paper can be used to design laser radiation detectors. Due to the condition $j \sim I_0$ stemmed from Eq. (35) such detectors could be apparently used to determine energy characteristics of a laser pulse, for example, intensity of laser pulses.

Experimental research of the drag current effect for charge carriers in low-dimensional semiconductive systems, in particular, for the drag current effect with the impurity centers participation, is not known for us. We note that the δ-alloyage technology (see, e.g., review [19]) is able, apparently, to provide necessary set up for such an experimental problem. It should be also noted that the predicted high sensitivity of the drag current effect to the energy spectrum, to the charge carriers momentum relaxation, and to optical transition types is of great interest within the framework of the semiconductive quantum structures fundamental physics.

As an example, let us estimate sensitivity $G$ of the photo-detector based on the one-electron drag current effect in a longitudinal magnetic field. In accordance with [20], $G = V/W$, where $V$ is the electromotive force value, $W = I_0 \cdot \hbar\nu \cdot S$ is intensity of the radiation, $\nu$ is frequency of the radiation, and $S$ is the light beam transversal section area.

In the "idle" regime, the value of $V$ can be found from equality condition in the given direction of the drag current to the corresponding conductance current. As the result, we obtain

$$G = \frac{V}{W} = \frac{S\rho_0}{S_0 W}\int_0^{L_{QW}} j\,dz \approx \frac{\rho_0 L_{QW} j}{I_0 S_0 \hbar\nu}, \qquad (36)$$

where $S_0$ is QW transversal section area and $\rho_0$ is QW material specific resistance. Considering the semiconductive structure consisting of single QW based on InSb and using the numerical values, $S_0 \sim L^2 \cong 1.8 \times 10^{-11}\ cm^2$, $I_0 \times \hbar\nu = 10^{10}\ Watt/m^2$, $L_{QW} \cong 7.2 \times 10^3\ nm$, $B = 10\ Ò$ , and $\rho_0 \sim 10^{-4}$ *ohm·m* in Eq. (36), for $\lambda \cong 6$ *mkm* we obtain the following estimation $G \approx 6.8 \times 10^{-2}$ *V/watt* . Hence, the one-dimensional electrons drag current effect in longitudinal magnetic field, is quite accessible for experimental study.

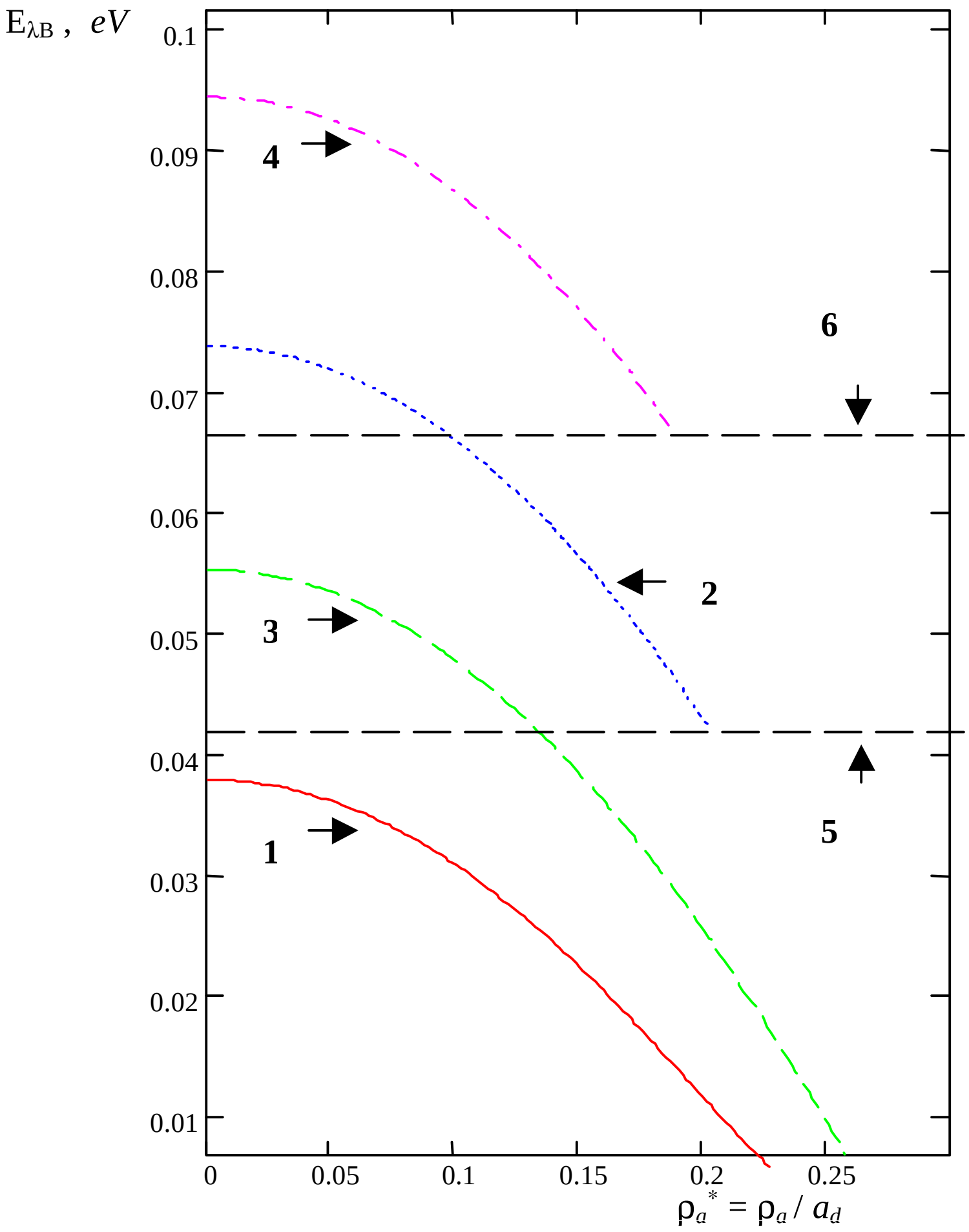

Fig. 1. The D(-)-center binding energy dependence, (under $2L = 71.6\ nm$ , $U_0 = 0.2\ eV$ ), from polar IC radius $\rho_a^* = \rho_a / a_d$ (in Bohr units) for various values of $B$ ; (curves 1 and 3 corresponds to the case $E_{\lambda_B}^{(0)} > 0$ , curves 2 and 4 correspond to the case $E_{\lambda_B}^{(0)} < 0$ ); the energy levels positions in the QW ground state for $B = 0\ T$ and $B = 12\ T$ are depicted by the dashed curves 5 and 6, correspondingly. **1** - $|E_i| = 5 \times 10^{-3}\ eV$ ; $B = 0\ T$ ; **2** - $|E_i| = 3.5 \times 10^{-2}\ eV$ ; $B = 0\ T$ ; **3** - $|E_i| = 5 \times 10^{-3}\ eV$ ; $B = 12\ T$ ; **4** - $|E_i| = 3.5 \times 10^{-3}\ eV$ ; $B = 12\ T$ .

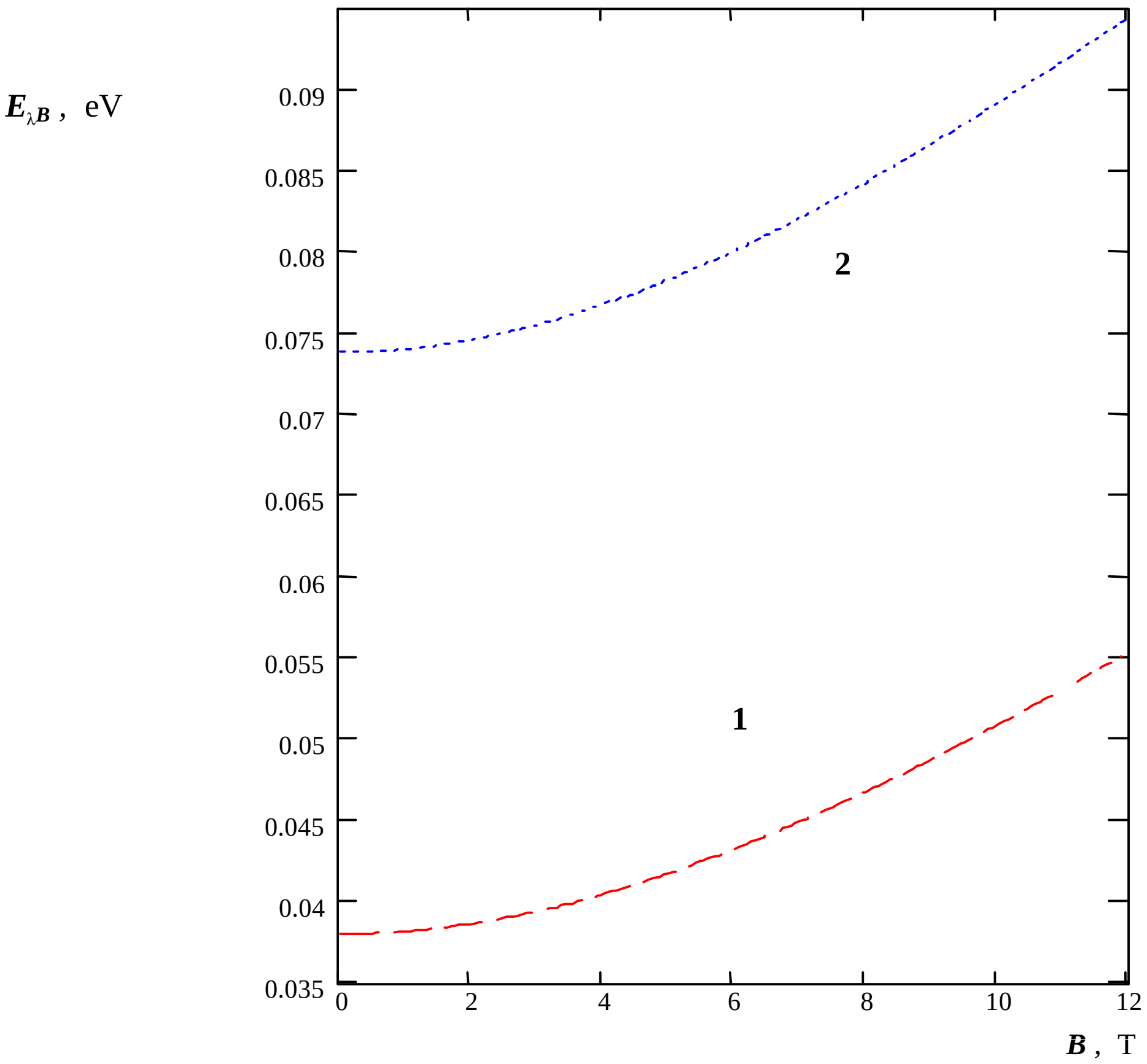

Fig. 2. The D$^{(-)}$-center (which is localized at the point $\vec{R}_a = (0,0,z_a)$; $2L = 71.6$ $nm$ , $U_0 = 0.2$ $eV$ ) binding energy dependence from $B$ values; (curve 1 corresponds to the case $E^{(0)}_{\lambda_B} > 0$ , curve 2 corresponds to the case $E^{(0)}_{\lambda_B} < 0$ ); **1** - $|E_i| = 5 \times 10^{-3}$ $eV$; **2** - $|E_i| = 3.5 \times 10^{-2}$ $eV$.

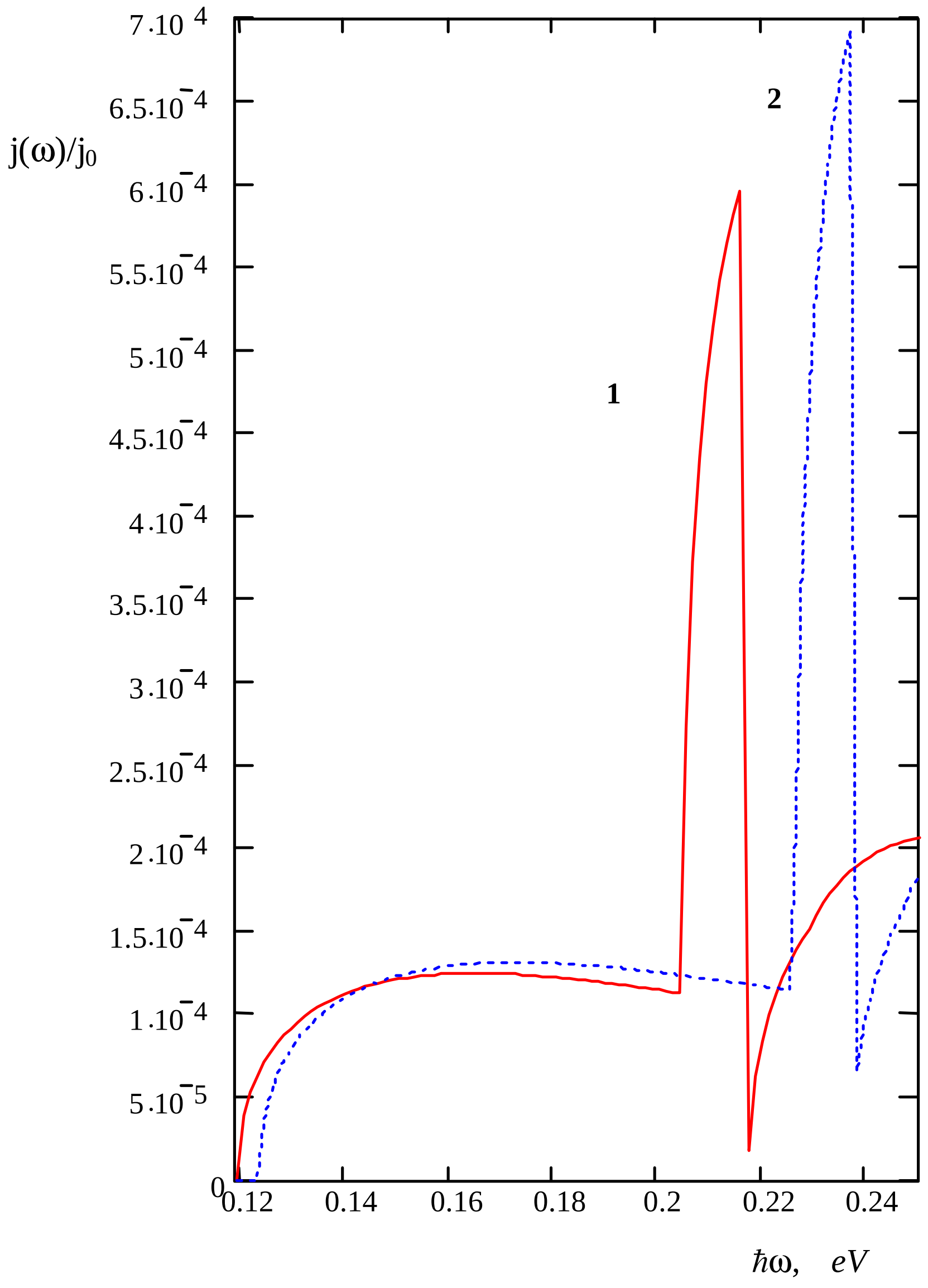

Fig. 3. The drag current density $j(\omega)/j_0$ spectral dependence (in relative units), at $|E_i|$ = $5.5 \times 10^{-2}$ eV ; $n_\lambda$ = $1.4 \times 10^5$ cm$^{-1}$; $2L$ = 71.6 nm ; $U_0$ = 0.2 eV; $n_e = 1.36 \times 10^{16}$ $cm^{-3}$ ; $n_i = 2.7 \times 10^{15}$ $cm^{-3}$ ; $\lambda_s$ = 28.6 $nm$ ; $T \approx 7$ $K$ , for different values of $B$ . Curve **1**: $B$ = 10 $T$ ; Curve **2**: $B$ = 12 $T$ .